\documentclass{article}
\usepackage{graphicx}
\usepackage{amsmath}
\usepackage{subfigure}
\usepackage{rsfso}
\usepackage{breqn}

\begin{document}

\title{\textbf{Quantum decoherence in Microtubules
}}


\author{\textit{Kaushik Naskar}  \\ Taki Government College,
	Taki-743429, Hasnabad, \\
	North 24 Parganas, West Bengal, India.    \\   \and
        \textit{Parthasarathi Joarder}  \\Sister Nivedita University ,
        DG 1/2 New Town Action Area 1, \\Kolkata - 700156, West Bengal, India. \\
}


           


\maketitle

\begin{abstract}
Not all activities in living creatures can be explained by classical dynamics. Application of quantum physics in biology helps to study the unexplained phenomena in cells. More detailed research work is needed rather than rejecting the concept of intervention of quantum physics in biology. Here we have used some concepts introduced by Hameroff, Penrose \cite{hamer} and some quantum models to show the quantum decoherence in neurons. Assuming a quantum superposition of dimers in microtubules we have separately presented two types of interaction with its environment. For interaction with bosonic environment we have shown that the decoherence time scale depends on a constant factor which depends on the interaction coefficients and amplitude of spectral density. For interaction with spin environment we have pointed out one case where the coherent superposition state of dimer is strong enough to survive against the environmental induced decoherence. 
\end{abstract}

\section{Introduction}
\label{intro}

Quantum biology focuses on phenomena in living creatures that cannot be fully explained by the classical physics. Among many examples, photosynthesis and magnetoreception in migratory birds are two areas where the explanations according to quantum theory are undeniable. Another intriguing and little-known topic is Consciousness. And the related processes in neurons are among the murky areas of biology where more detailed scientific works are needed.

\indent Classical dynamics have not yet been able to fully describe the concept of consciousness. Some research has been done to comprehend consciousness utilising quantum physics. Hameroff and Penrose\cite{penrose}\cite{hamer} have introduced quantum superposition  in microtubules and used quantum gravity to calculate the mechanism of self-collapse which they termed ``orchestrated objective reduction"(``Orch OR"). This idea and the associated model was criticized by several researchers\cite{tegmark}\cite{rosa}. Nevertheless, it provided a new avenue for consciousness-related brain research. Here in Section 1, we briefly describe the structure of microtubules in neurons and the introduction of quantum superposition in microtubules. The following sections discuss the two quantum models in microtubules and finally we conclude with the implications of the quantum model for neurons in the brain.

\section{Microtubules in neurones}
\label{mn}

\begin{figure}
\centering
\includegraphics[width=0.5\textwidth]{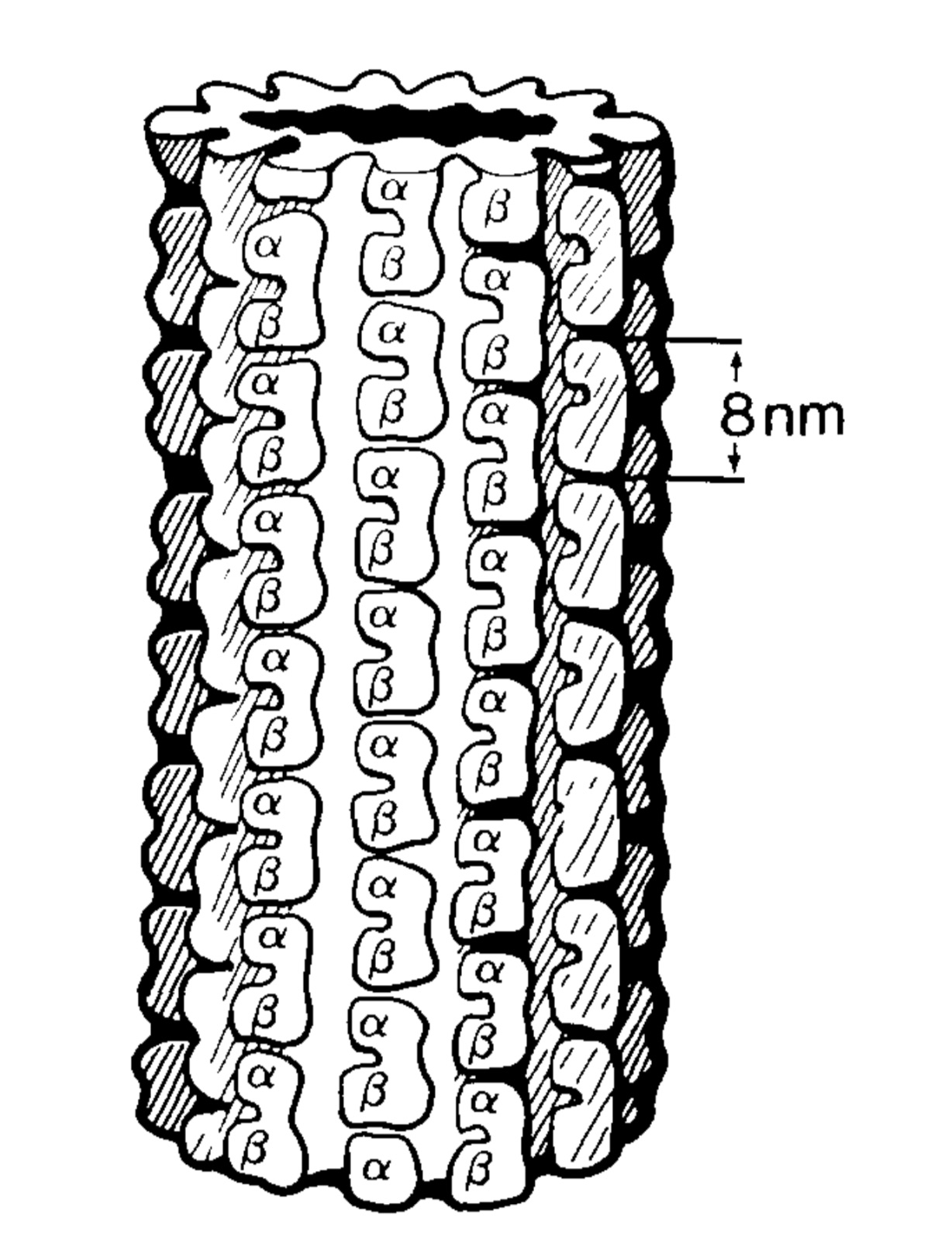}
\caption{Structure of microtubule. The picture is taken from Hameroff and Penrose\cite{hamer}. Diameter of the microtubule is 25nm. $\alpha$ and $\beta$ are two subunits or monomers which create a dimer of length 8nm.}
\label{mt}     
\end{figure}
The fundamental components of the brain are neurons, also known as nerve cells. The main functional component for preserving the structure of neurons is the cytoskeleton. Protein polymers in the form of network serve as cytoskeleton in the neurons. Hollow cylindrical parts of this network are microtubules. Fig. \ref{mt} shows the structure of microtubules. The diameter of microtubule is 25 nanometers(nm). Along its longitude, the wall of the microtubule is made up of 13 protein fibres or protofilaments. These protofilaments are called tubules which are consisted by dimers of length 8nm. These dimers are known as tubulins. Each tubulin is made up of two subunits, $\alpha$ tubulin and $\beta$ tubulin. 

\indent Each dimer has two hydrophobic pockets and one delocalized electron. One pocket(let's call it $P_\alpha$) is slightly towards the $\alpha$ tubulin and the other(let's call it $P_\beta$) towards the $\beta$ tubulin. The delocalized electron can choose either of the two hydrophobic pockets to localize. According to the localization of the electron two states are defined; $\alpha$ state(if it localizes at $P_\alpha$) and $\beta$ state(if it localizes at $P_\beta$). Considering the theories of Fr$\ddot{o}$hlich\cite{frohlich1}\cite{frohlich2}\cite{frohlich3}, Hameroff and Penrose\cite{hamer} showed that due to dipole oscillation in quantum level the two states, $\alpha$ and $\beta$, become a coherent superposition(say, $\alpha \beta$ state).

\section{Quantum Decoherence in tubulin}
\label{qd}
Quantum decoherence is the phenomenon through which a coherent superposition or entanglement of quantum states is destroyed by the interaction with the environment. Due to this inevitable phenomenon, the coherent state of the tubulin, $\alpha \beta$ state, interacts with its environment inside neuron. According to the particles of the environment the nature of interaction changes. The following two subsections describe two types of environment and interactions\cite{leggett}-\cite{schl}.
\subsection{\textbf{Bosonic environment}}
In this subsection following some well-known methods and procedures\cite{leggett}\cite{reina}\cite{breuer}\cite{naskar} we describe the interaction with an environment which is composed of bosonic particles(like hydrogen, oxygen atom etc.). The $\alpha \beta$ state can be regarded as a coherent state of spin-$\frac{1}{2}$ particle. Suppose the state evolves according to the self Hamiltonian 
\begin{equation}
\hat{H}_{\alpha\beta} = \frac{1}{2}\omega \hat{\sigma}_{zs}
\end{equation}
and interacts with the environment according to the interaction Hamiltonian
\begin{equation}
\hat{H}_{\alpha\beta E}=\hat{\sigma}_{zs} \otimes \sum_i a_i \hat{x}_i
\label{inter}
\end{equation}
then the total Hamiltonian will be
\begin{equation}
\hat{H}_{tot} = \hat{H}_{\alpha\beta} +\hat{H}_{\alpha\beta E} + \hat{H}_E
\label{hamil}
\end{equation}
where $\omega$ is the energy difference between $\alpha$ and $\beta$ states, $\hat{\sigma}_{z}$ represents the Pauli matrix or spin observable along z-axis and the subscription $s$ is used to denote the system, $\hat{x}_i$'s are the positions of the particles of the environment and $a_i$'s are the corresponding coupling coefficients with $\alpha\beta$ state, the environmental self-Hamiltonian is denoted by $\hat{H}_E$ which is nothing but the collection of Hamiltonians of harmonic oscillators. 

\indent Let the environment inside neuron is in thermal equilibrium at the normal temperature(say, $T$) of human body.  Due to the interaction Hamiltonian (\ref{inter}) the $\alpha\beta$ state gets entangled with the environmental states and the composite state of the superposed tubulin and the environment evolves according to the total Hamiltonian (\ref{hamil}) with time. Calculations of tedious mathematical procedures show that the diagonal elements of the reduced density matrix of the $\alpha\beta$ state, $\hat{\rho}_{\alpha\beta}(t)$ do not change with time  $t$. But here the off-diagonal elements of $\hat{\rho}_{\alpha\beta}(t)$, which change with time, are the most important things. These terms contain a factor like $\mathcal{D}(t)=e^{-\gamma (t)}$, which is called \textit{Decoherence factor} and it depicts that the off-diagonal elements decrease with time. Using Ohmic spectral density for dense environment, $\gamma (t)$ can be shown to have the following form 
\begin{equation}
\gamma(t)=C_0 \int_{0} ^{\infty}\frac{e^{-\frac{\omega}{\Omega}}}{\omega}coth\left(\frac{\omega}{2k_B T}\right)(1-cos\omega t) d\omega
\label{gmm}
\end{equation}
where $k_B$ is the well-known Boltzmann constant, $\Omega$ is an upper cutoff frequency for the spectral density and $C_0$ is another constant which will be discussed later. If we consider that the interaction time between the system(i.e. $\alpha\beta$ state) and the environment is quite larger than the thermal fluctuation time of the environment then equation (\ref{gmm}) can be simplified to
\begin{equation}
\gamma(t)\approx \pi C_0 k_B T t.
\end{equation}
This expression implies that the $\alpha\beta$ state, which was initially a coherence superposition state, decoheres after a time scale of $1/\pi C_0 k_B T$ and it is termed as decoherence time scale $\tau_d$.  Taking the exact values of the constants, $\tau_d$ can be calculated and we can find $\tau_d=1.60485 \times 10^{-21} C_0$ sec. The value of $C_0$ depends on the strength of interaction. The type and density of an environment define the spectral density of the environment which takes a vital role in calculation of $\gamma(t)$ and some factors of spectral density are included in $C_0$. Thus $C_0$ also depends on the spectral density of the environment. Now to find the exact value of $\tau_d$ finding the proper value of $C_0$ for the particular environment inside the nerve cells is important as well as difficult. Finding the proper value of $C_0$ is our future proposed work.

\subsection{\textbf{Spin environment}}

Here following the procedures of Cucchietti, Paz and Zurek\cite{cucc} we discuss the interaction with spin environment, an environment of spin-$\frac{1}{2}$ particles. For simplicity we assume a strong interaction between the system($\alpha\beta $ state) and the environment. The corresponding interaction Hamiltonian may be written as $\hat{H}_{int}=\frac{1}{2}\hat{\sigma}_{zs}\otimes\sum_i b_i \hat{\sigma}_{zi}$, where $b_i$'s are the coupling coefficients bteween the system and the $i^{th}$ spin of the environment, $\hat{\sigma}_{zs}$ is the same as in equation (\ref{inter}) and the $i$ in place of the subscription $s$ is denoted for the $i^{th}$ spin of the environment. The total Hamiltonian for the system and the environment is given by
\begin{equation}
	\hat{H}_{tot}= \hat{H}_{tun} + \hat{H}_{int}
\end{equation}

where the tunneling Hamiltonian $\hat{H}_{tun}= -\frac{1}{2} E_t \hat{\sigma}_{xs}$ is included in the total Hamiltonian for more generic case, $E_t$ is the tunneling coefficient and $\hat{\sigma}_{xs}$ is another Pauli spin matrix for the system along x-axis.
As $\hat{H}_{int}$ is a diagonal matrix in the eigen basis of $\hat{\sigma}_z$ thus it can be written as
\begin{equation}
\hat{H}_{int} = \frac{1}{2}	E_{v}\hat{\sigma}_{zs}
\end{equation}
where $E_{v} = \sum_{i} S_i b_i$ is the eigen value of the environmental part of the interaction Hamiltonian and $S_i$ can have values either -1 or +1 according to the corresponding states of the environment. The initial combined density density operator $\hat{\rho}_{SE}(t=0)$ of the system and the environment evolves according to the evolution operator
\begin{equation}
	\hat{\mathcal{U}}_{SE}(t)=exp[-\frac{i}{2}(-E_t\hat{\sigma}_{xs}+E_{v}\hat{\sigma}_{zs})t]
\end{equation} 

and after time $t$ it takes the form as
\begin{dmath}
	\hat{\rho}_{SE}(t)=\left[\hat{\boldsymbol{I}} cos\left(\sqrt{E_{v}^2 + \frac{E_t ^2}{4}}t\right) - i \left(E_{v}\hat{\sigma}_{zs}-\frac{1}{2}E_t\hat{\sigma}_{xs}\right)\frac{sin\left(\sqrt{E_{v}^2 + \frac{E_t ^2}{4}}t\right)}{\sqrt{E_{v}^2 + \frac{E_t ^2}{4}}}\right] \hat{\rho}_{SE}(0) \left[\hat{\boldsymbol{I}} cos\left(\sqrt{E_{v}^2 + \frac{E_t ^2}{4}}t\right) - i \left(E_{v}\hat{\sigma_{zs}}-\frac{1}{2}E_t\hat{\sigma}_{xs}\right)\frac{sin\left(\sqrt{E_{v}^2 + \frac{E_t ^2}{4}}t\right)}{\sqrt{E_{v}^2 + \frac{E_t ^2}{4}}}\right]^\dagger
\end{dmath}

\begin{figure}
	\subfigure[]{\includegraphics[width=0.5\textwidth]{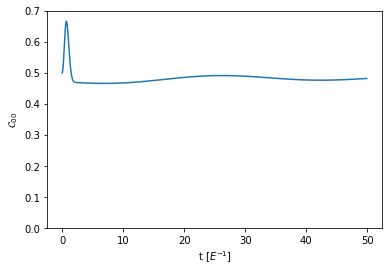}{\label{fig2a}}}
	\subfigure[]{\includegraphics[width=0.5\textwidth]{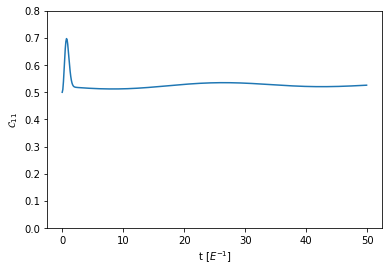}{\label{fig2b}}}
	\subfigure[]{\includegraphics[width=0.5\textwidth]{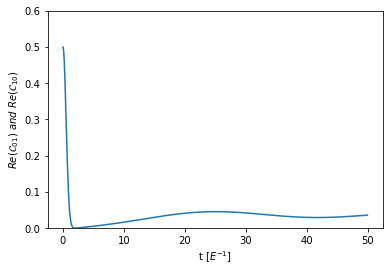}{\label{fig2c}}}
	\subfigure[]{\includegraphics[width=0.5\textwidth]{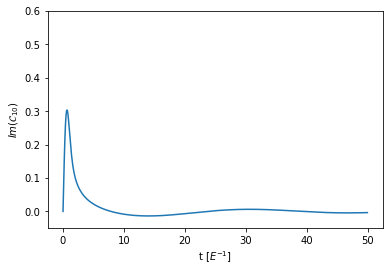}{\label{fig2d}}}
	\caption{Graphical representation of the coefficients $\mathcal{C}_{ij}$ for $E_t/d=0.1$. Fig. \ref{fig2a} and \ref{fig2b} show that for $t>1.5[E^{-1}]$  the value of the diagonal elements of the reduced density matrix of the system approach to a finite number. But Fig. \ref{fig2c} and Fig. \ref{fig2d} depict that the values of the real and imaginary parts of the off-diagonal elements separately approach to zero for the same time scale.}
	\label{coff1}      
\end{figure}

Now considering a continuous Gaussian spectral density for large and dense medium we get the reduced density operator for the system as
\begin{equation}
	\hat{\rho}_S (t) = \left[\mathcal{C}_{00}|0\rangle\langle 0|+\mathcal{C}_{01}|0\rangle\langle 1|+\mathcal{C}_{10}|1\rangle\langle 0|+\mathcal{C}_{11}|1\rangle\langle 1|\right]
\end{equation}
where
\begin{eqnarray*}
&\mathcal{C}_{00}&=\frac{1}{2\sqrt{2\pi d^2}}\int[cos^2(Et) + \frac{sin^2(Et)}{E^2}(E_{v}-E_t)^2]exp\left(-\frac{E_{v} ^2}{2d^2}\right) dE_{v}, \\
&\mathcal{C}_{01}&=\mathcal{C}_{10}^\dagger=\frac{1}{2\sqrt{2\pi d^2}}\int[cos^2(Et) + \frac{sin^2(Et)}{E^2}(\frac{E_t ^2}{4}-E_{v} ^2) \\&&~~~~~~~~~~~- i2E_{v} \frac{sin(Et)}{E}cos(Et)]exp\left(-\frac{E_{v} ^2}{2d^2}\right) dE_{v}, \\
&\mathcal{C}_{11}&=\frac{1}{2\sqrt{2\pi d^2}}\int[cos^2(Et) + \frac{sin^2(Et)}{E^2}(E_{v}+E_t)^2]exp\left(-\frac{E_{v} ^2}{2d^2}\right) dE_{v},
\end{eqnarray*}

\begin{figure}
	\subfigure[]{\includegraphics[width=0.5\textwidth]{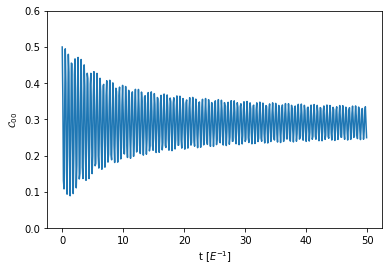}{\label{fig3a}}}
	\subfigure[For $t\gg \omega_c ^{-1}$ but $t\ll \beta$]{\includegraphics[width=0.5\textwidth]{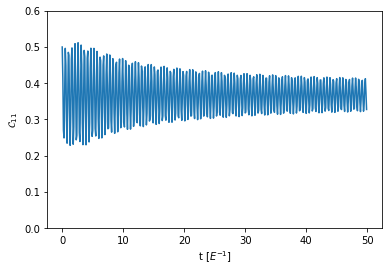}{\label{fig3b}}}
	\subfigure[For $t\gg \beta$]{\includegraphics[width=0.5\textwidth]{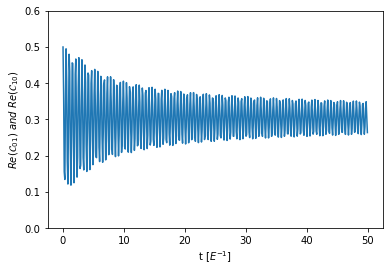}{\label{fig3c}}}
	\subfigure[For hkh]{\includegraphics[width=0.5\textwidth]{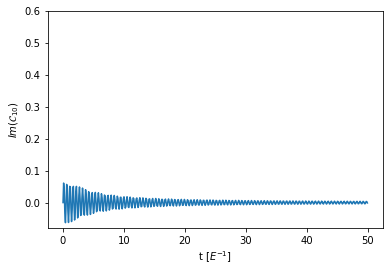}{\label{fig3d}}}
	\caption{Graphical representation of the coefficients $\mathcal{C}_{ij}$ for $E_t/d=6$. Fig. \ref{fig3a} and Fig. \ref{fig3b} represents the diagonal elements of $\hat{\rho}_S(t)$. These two figures show that as time passes the value of the corresponding coefficients approach to a non-zero finite number. Similar phenomenon is observed for the real parts of the off-diagonal elements of $\hat{\rho}_S(t)$ as we can see in Fig. \ref{fig3c}. In Fig. \ref{fig3d} we see that the imaginary parts of the off-diagonal elements of $\hat{\rho}_S(t)$ tends to zero.}
	\label{coff2}      
\end{figure}

$E=\sqrt{E_{v}^2+\frac{E_t ^2}{4}}$ and $d$ is tha standard deviation of the assumed spectral density.  
\begin{figure}
	\subfigure[For $E_t/d=0.2$]{\includegraphics[width=0.5\textwidth]{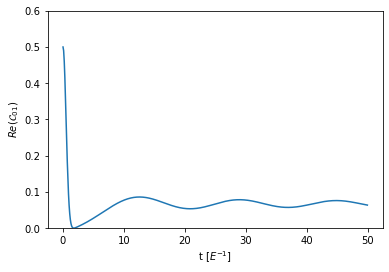}{\label{fig4a}}}
	\subfigure[For $E_t/d=0.5$]{\includegraphics[width=0.5\textwidth]{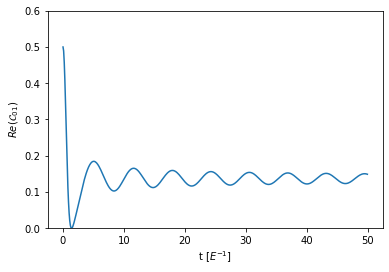}{\label{fig4b}}}
	\subfigure[For $E_t/d=1.0$]{\includegraphics[width=0.5\textwidth]{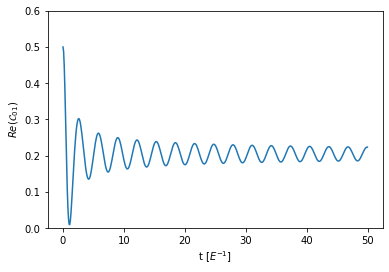}{\label{fig4c}}}
	\subfigure[For $E_t/d=2.0$]{\includegraphics[width=0.5\textwidth]{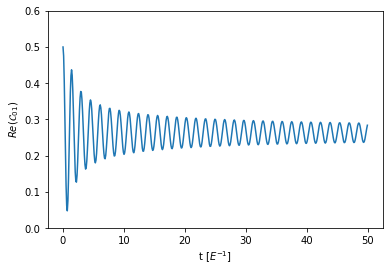}{\label{fig4d}}}
	\caption{Graphical representation of $Re(\mathcal{C}_{01})$ for different values of $E_t/d$. It shows that $Re(\mathcal{C}_{01})$ tends to a larger number as $E_t/d$ increases.}
	\label{coff3}      
\end{figure}

Fig. \ref{coff1} shows the time dependence of above coefficients $\mathcal{C}_{ij}$ when $E_t/d$ is quite less than 1. We have taken $E_t/d = 0.1$. Fig. \ref{fig2a} and Fig. \ref{fig2b} show that the values of $\mathcal{C}_{00}$ and $\mathcal{C}_{11}$ approach towards a non-zero number with increase in time. These are the diagonal elements of the reduced density matrix of the system($\alpha\beta$ state). Fig. \ref{fig2c} shows the variation of real parts of $\mathcal{C}_{01}$ and $\mathcal{C}_{10}$. It falls rapidly with time and after $t\sim 1.5[E^{-1}]$ it becomes almost zero. The imaginary part of $\mathcal{C}_{10}$ is shown in Fig. \ref{fig2d}. Imaginary part of $\mathcal{C}_{01}$ have the same mathematical expression with a -ve sign. Therefore, the graphical representation of it has the same nature in the fourth quadrant. So the imaginary parts of $\mathcal{C}_{01}$ and $\mathcal{C}_{10}$ become almost zero along with a negligible amplitude as time increases. So, Fig. \ref{coff1} depicts that after little time, the off-diagonal elements of  $\hat{\rho}_S (t)$ vanish but the diagonal elements survive inspite of the environment induced decoherence. This implies that the $\alpha\beta$ state will collapse to either $\alpha$ or $\beta$ state.

For $E_t/d$ is quite larger than 1, the coefficients $\mathcal{C}_{ij}$ are shown in Fig. \ref{coff2}. We have assumed $ E_t/d= 6$. Time dependence of $\mathcal{C}_{00}$, $\mathcal{C}_{11}$, $Re(\mathcal{C}_{01})$ or $Re(\mathcal{C}_{10})$ and $Im(\mathcal{C}_{10})$ are represented in Fig. \ref{fig3a}, Fig. \ref{fig3b}, Fig. \ref{fig3c} and Fig. \ref{fig3d} respectively. Among all these 4 graphs only the imaginary part of $\mathcal{C}_{10}$, i.e. $Im(\mathcal{C}_{10})$, tends to zero with increase in time. We expect similar nature for $Im(\mathcal{C}_{01})$. But $\mathcal{C}_{00}$, $\mathcal{C}_{11}$, $Re(\mathcal{C}_{01})$ and $Re(\mathcal{C}_{10})$ approach towards some non-zero finite numbers. This imply that no element of the reduced density matrix of the system i.e., $\hat{\rho}_{S}(t)$ vanishes. Therefore the system is still in a coherent superposition state in the eigen basis of $\hat{\sigma}_z$. 

Fig. \ref{coff3} shows the time dependence of $Re(\mathcal{C}_{01})$ for different values of $E_t/d$. It clearly shows that the value of $Re(\mathcal{C}_{01})$ approaches to a higher number with increase in the value of $E_t/d$. Similar conclusions can be drawn for $Re(\mathcal{C}_{10})$. Therefore, the effect of decoherence reduces with increase in the value of the ratio $E_t/d$.

\section{Conclusions}
Concept of consciousness and function of nervous system  are not clearly understood by classical dynamics. Microtubules are one of the most vital units in neurones of nervous system. Considering the smallest part of microtubules, i.e. tubulins, in a quantum coherent superposition state($\alpha\beta$ state) we assumed the interaction of tubulins or dimers with two different types of environment. For a bosonic environment we obtain that the superposition state decohere at a time scale of $1.60485 \times 10^{-21} C_0$ sec. $C_0$ depends on the coupling strength of the interaction and the spectral density. Finding the explicit value of $C_0$ for real environment inside neuron is very important for finding the actual time of decoherence. This is our proposed future work. For a spin environment there is another important ratio $E_t/d$ which decides the effect of decoherence. If the value of $E_t/d \ll 1$ then decoherence affects the quantum state of the dimer more and the dimer becomes either $\alpha$ tubulin or $\beta$ tubulin. But if $E_t/d \gg 1$, decoherence does not greatly affect the superposition state of the dimer. Consequently the state remains in the superposition state in the eigen basis of $\hat{\sigma}_z$ until measurement.

\end{document}